\documentstyle[twoside,amssymb,12pt]{article}

\newcommand{\nc}{\newcommand}
\nc{\la}{\lambda} \nc{\alf}{\alpha}
\nc{\tht}{\theta}  \nc{\be}{\beta}  \nc{\eps}{\epsilon} \nc{\ze}{\zeta}
\nc{\ga}{\gamma}  \nc{\De}{\Delta}  \nc{\Ga}{\Gamma}  \nc{\vphi}{\varphi}
\nc{\de}{\delta} \nc{\si}{\sigma}  \nc{\ka}{\kappa}   \nc{\Si}{\Sigma}
\nc{\om}{\omega}  \nc{\qq}{\quad\quad}                \nc{\Om}{\Omega}
\nc{\nf}{\infty}   \nc{\dl}{\mathop{\smash{\cal L}}}  \nc{\black}{\rule{3mm}{3mm}}
\nc{\ra}{\rightarrow}  \nc{\ol}{\overline}  \nc{\und}{\underline}
\nc{\beq}{\begin{equation}}  \nc{\pt}{\partial}  \nc{\nin}{\noindent}
\nc{\eeq}{\end{equation}}
\nc{\beqa}{\begin{eqnarray}}  \nc{\dst}{\displaystyle}
\nc{\eeqa}{\end{eqnarray}} \nc{\nnb}{\nonumber}
\nc{\bs}{\backslash}        \nc{\mb}{\mathbb}

\newcounter{muni}
\newenvironment{remunerate}{\begin{list}{{\rm \arabic{muni}.}}
{\usecounter{muni}
\setlength{\leftmargin}{0pt}\setlength{\itemindent}{38pt}}}{\end{list}}

\nc{\brm}{\begin{remunerate}}   \nc{\erm}{\end{remunerate}}

\newtheorem{nlem}{Lemma} \newtheorem{nth}{Theorem}
\nc{\stg}{\mathop{\smash{*}}}
\nc{\st}{\mathop{\smash{\delta}}}
\nc{\barr}{\begin{array}}   \nc{\earr}{\end{array}}   \nc{\dg}{\dagger}
\nc{\mtvb}{\mathversion{bold}}   \nc{\mtvn}{\mathversion{normal}}
\nc{\ti}{\tilde}  \nc{\wti}{\widetilde} \nc{\wh}{\widehat}

\begin{document}

\begin{titlepage}
\begin{flushright}
5 may 2013
\end{flushright} 
\vskip 2.0truecm
\centerline{\large \bf REPLY TO PROFESSOR YEHIA COMMENTS}
\vskip 1.0truecm 
\centerline{\bf Galliano Valent${}^{*}$}
\vskip 2.0truecm 

\centerline{${}^{*}$\it Laboratoire de Physique Th\'eorique et des
Hautes Energies}
\centerline{\it Unit\'e associ\'ee au CNRS UMR 7589}
\centerline{\it 2 Place Jussieu, 75251 Paris Cedex 05, France} 
%\vskip 3truecm

%\end{titlepage} 

\section{Reply and explanations} 
In a recent Comment \cite{Ye1} Pr Yehia claims that two of my articles \cite{Va1}, \cite{Va2} 
do not quote properly his previous articles and in fact contain nothing new with respect to his results. 
We will show that these statements are not true and try to explain what were the aims of these two articles and what were the relevant new results.

\subsection{The first article}
This is article \cite{Va1} which deals with a class of integrable systems with a cubic first integral. In this article 3 pages deal with a derivation of the explicit closed form of the integrable system  (which is described by a finite number of parameters) while the remaining 15 pages deal with the 
global structure (for which parameters is the system defined on a manifold and what is this manifold).

The paper appeared and Professor Yehia wrote to the review to mention his two articles \cite{Ye2}, 
\cite {Ye3} where he had obtained the explicit local form. It was obvious that these works were the 
first to solve the local problem but contained no global analysis. So an erratum was published in CMP 
\cite{Va3}, where the two articles by Pr Yehia are properly quoted: he had solved the local problems 
but not the global ones.

It is true that his article \cite{Ye3} was quoted in the published form of Dullin and Matveev \cite{dm}, but very unfortunately I had used only their arXiv preprint in which he was not quoted.

So this point is now clear: my article contains new (global) results which are not exhausted by the local analyses of Pr Yehia and his work was quoted properly in the erratum.

\subsection{Second article}
This is the preprint \cite{Va2} which deals with a class of integrable systems with a quartic first integral. In this article 3 pages deal with a derivation of the explicit closed form of the integrable system (which is again described by a finite number of parameters) while the remaining 28 pages deal with the global problems.

I was aware of a work by Hadeler and Selivanova \cite{hs} where the local structure had already been obtained. This article, which  appeared in {\sl Regular and Chaotic Dynamics} in 1999, contains also 
global results for some models on the manifold $S^2$. One can check that this article is properly 
quoted in the Remarks at the bottom of page 5 of my article.

Now it happens that Pr Yehia has studied integrable systems (locally, as usual) with quartic integrals in his article \cite{Ye4} but which appeared in 2006, {\em so he cannot claim to have priority}. It is 
amusing to  notice that the prior work by Hadeler and Selivanova is not quoted in his article! 
Certainly he has obtained even more general integrable systems but which are not of concern to me. Nevertheless I quoted his article in my conclusion since it would deserve certainly more work on the global issues.

I hope that it is now clear: my article contains new global results on these integrable models with a quartic integral, the local form of which was first obtained by Hadeler and Selivanova and the reference \cite{Ye4} by Pr Yehia is properly quoted.

\section{Conclusion}
I am happy to observe that these two articles of mine had at least one reader: Pr Yehia. More 
seriously I think that the proper quotation of prior articles is becoming a very difficult problem, 
due to the enormous amount of results buried in the reviews. The best to do is to always put the articles on arXiv: in such a way if some reference by X is missing no doubt that X can send an e-mail to the author and this would be sufficient to settle the problem.

\end{titlepage} 

\end{document}